\documentclass{elsart}

\usepackage{graphicx}
\usepackage{natbib}
\usepackage{amssymb,amsmath}

\newcommand{\p}{_{\text{p}}}
\newcommand{\eff}{_{\text{eff}}}
\newcommand{\sub}{_{\text{s}}}
\newcommand{\sstar}{_{\star}}
\newcommand{\sun}{_\odot}
\newcommand{\MM}{\mathcal{M}}
\newcommand{\E}{_{\text{E}}}
\newcommand{\ts}{\tau_{\text{sync}}}

\begin{document}
\begin{frontmatter}

\title{On the protection of extrasolar Earth-like planets 
around K/M stars against galactic cosmic rays}

\author[LESIA,ASTRON]{J.--M. Grie{\ss}meier\corauthref{cor}},
\corauth[cor]{Corresponding author.}
\ead{griessmeier@astron.nl}
\author[TUBS]{A. Stadelmann},
\author[DLR,TUBE]{J. L. Grenfell},
\author[IWF]{H. Lammer},
\author[TUBS]{U. Motschmann}

\address[LESIA]{Laboratoire d'Etudes Spatiales et d'Instrumentation en Astrophysique
(LESIA), Observatoire de Paris, CNRS, UPMC, Universit\'{e} Paris Diderot; 5 Place Jules Janssen,
92190 Meudon, France}
\address[ASTRON]{Netherlands Institute for Radio Astronomy, 
Postbus 2, 7990 AA, Dwingeloo, The Netherlands}
\address[TUBS]{Technische Universit\"{a}t Braunschweig, Mendelssohnstra{\ss}e 3, 
38106 Braunschweig, Germany}
\address[DLR]{Institut f\"{u}r Planetenforschung,
Deutsches Zentrum f\"{u}r Luft- und Raumfahrt (DLR), Rutherford Str. 2, 12489 Berlin, Germany}
\address[TUBE]{Zentrum f\"{u}r Astromomie und Astrophysik, Technische 
Universit\"{a}t Berlin (TUB), Hardenbergstr. 36, 10623 Berlin, Germany}
\address[IWF]{Space Research Institute, Austrian Academy of Sciences, Schmiedlstr. 6, A-8042 Graz, Austria}


\begin{abstract}

Previous studies have shown that extrasolar Earth-like planets in close-in habitable zones 
around M-stars are weakly protected against galactic cosmic rays (GCRs), leading
to a strongly increased particle flux to the top of the planetary atmosphere. Two main effects were held 
responsible for the weak shielding of such an exoplanet: (a) For a close-in planet, the planetary 
magnetic moment is strongly reduced by tidal locking. 
Therefore, such a close-in extrasolar planet is not protected by an extended magnetosphere.
(b) The small orbital distance of the planet exposes it to a much denser stellar wind than that prevailing 
at larger orbital distances. This dense stellar wind leads to additional compression of the 
magnetosphere, which can further reduce the shielding efficiency against GCRs.
In this work, we analyse and compare the effect of (a) and (b), showing that 
the stellar wind variation with orbital distance has little influence on the cosmic ray shielding. Instead, the weak shielding
of M star planets can be attributed to their small magnetic moment.
We further analyse how the planetary mass and composition influence the planetary magnetic 
moment, and thus modify the cosmic ray shielding efficiency.
We show that more massive planets are not necessarily better
protected against galactic cosmic rays, but that the planetary bulk composition can play an important
role.

\end{abstract}


\end{frontmatter}

\section{Introduction}

One of the many fascinating questions in the field of exoplanet studies is the search for 
habitable worlds.
Because of their relatively small mass, low luminosity, long lifetime and large abundance in the
galaxy, M dwarfs are sometimes suggested as prime targets in searches for terrestrial habitable 
planets \citep{Tarter07, Scalo07}. 
The detection of planets with $M\le 10 M\E$, i.e.~with a mass smaller than 10 terrestrial masses
\citep[the upper limit for ``super-Earths'' in the classification of][]{Valencia07b}
recently became possible with current advances in instrumentation and analysis. 
The first super-Earth detected was GJ 876d, a planet with $\sim\!\!7.5 M\E$ 
or\nolinebreak[4]biting an M star \citep{Rivera05}. 
Thereafter, other low-mass planets have been discovered: 
OGLE-2005-BLG-390Lb, a planet with $5.5 M\E$ in an orbit of 5 AU was found by microlensing 
\citep{Beaulieu06}, 
and HD 69830b, a planet with a minimum mass of $10.2 M\E$ in an orbit of 0.08 AU was detected by the radial
velocity technique \citep{Lovis06}.
More recently, two super-Earth planets were discovered around the M3 dwarf Gl 581, with minimum 
masses of 5.0 $M\E$ and 7.7 $M\E$, and semimajor axes of 0.07
and 0.25 AU \citep{Udry07}, 
and \citet{Ribas08} report the possible detection of a 5.0 $M\E$ planet
around the M-type star GJ 436.
In total, 6 planets with masses $\lesssim 10 M\E$ are known today, 5 of which are located in an 
orbit around an M star\footnote{Up-to-date numbers can be found at the Extrasolar Planets 
Encyclopaedia: {\tt http://www.exoplanet.eu}}. 
The least massive planet has a mass of 5.0 $M\E$.
While the number of super-Earth planets around M stars is currently still limited, 
the detection of cold debris disks around M stars seems to indicate that planets may be as
frequent for M stars as they are for F, G and K stars \citep{Lestrade06}. 

For M stars the habitable zone \citep[defined by the range of orbital distances
over which liquid water is possible on the planetary surface, see e.g.][]{Kasting93}
is much closer to the star (depending on stellar mass, but typically $\le$0.3 AU).  
Such close-in distances pose additional problems and constraints to habitability. 
For example,
there is the possibility that the atmosphere might collapse on the night side of the planet,
although it seems that this problem is less severe than previously assumed 
\citep[][]{Joshi97,Joshi03}.
Intense stellar flares are common, especially for young M stars, but the radiation can be
shielded by the planetary atmosphere \citep{Heath99}. Both direct UV radiation
\citep{Buccino07} and indirect UV radiation generated by energetic particles \citep{Smith04} 
can have important consequences for biogenic processes on M-star planets.
Another potential problem for M star planet habitability is that one expects a large number of 
coronal mass ejections (CMEs) on active M stars. 
The related interaction of the dense CME plasma flux with the 
atmosphere/magnetosphere environment of the exposed planets during the active 
stage of the stellar evolution could be strong enough to erode the atmosphere 
or the planetary water inventory via
non-thermal atmospheric loss processes  \citep{Khodachenko07AB, Lammer07AB}. 

The efficient loss of water from the planet could also lead to a breakdown of 
plate tectonics. If the majority of the planets' water is lost, the lithosphere 
will not be sufficiently deformable, and subduction of the crust cannot occur. 
The recycling of the crust through plate tectonics keeps the crust thin. If the 
crust is too thick, the lithospheric plate comprising the crust will be too 
buoyant to be subducted. A thin crust and a wet planet seem to be very important
for plate tectonics to operate. Plate tectonics, however, help to cool the 
interior efficiently which is required to keep an Earth-like thermally driven 
magnetic dynamo in operation over geological time spans \citep{Lammer08}.
Even in the case of M star planets with a magnetic dynamo, the magnetic moment 
is not likely to be large.
Terrestrial planets in the habitable zones of M-stars
can be expected to be tidally locked, which leads to a reduction in 
the planetary magnetic moment \citep{Griessmeier05a}.
For these reasons, one can assume that M-star planets within the habitable zones are only weakly protected by their magnetospheres, resulting in a higher cosmic ray exposure.
In addition, the small orbital distances of these planets expose it to much denser stellar winds 
than at larger orbital distances. The dense stellar winds lead to a further compression of the 
magnetospheres, which can in principle again reduce the shielding efficiency against GCRs.

In this work we extend the studies of \citet{Griessmeier05a} and \citet{GriessmeierESLAB05} 
by separately quantifying and comparing these two effects
(effect of enhanced stellar wind ram pressure 
due to the smaller orbital distance vs.~influence of a reduced planetary magnetic dipole moment).
We show that the 
stellar wind variation with orbital distance
has little influence on the shielding efficiency against
galactic cosmic rays. 
The weak shielding found for M star planets \citep{Griessmeier05a,GriessmeierESLAB05} thus has 
to be attributed to their reduced magnetic moment.
Given that the magnetic moment is the key parameter, we further analyse the influence of the 
planetary mass and its composition on the planetary magnetic moment, and thus on the cosmic ray
shielding efficiency.

This paper is organised as follows:
The planetary parameters governing the size of the magnetosphere are presented in Section
\ref{sec-planet}.
We show that planets within the habitable zone of
M stars are tidally locked, which leads to a reduced planetary magnetic dipole
moment (Section \ref{sec-magmoment}). The stellar wind parameters at close orbital distances are
discussed in Section \ref{sec-stellarwind}. 
The influence of tidal locking and stellar wind on the size of the planetary magnetosphere is
briefly mentioned in Section \ref{sec-magnetosphere}.
In Section \ref{sec-distance1}, we discuss  
the dependence of the cosmic ray 
flux \textit{outside} the planetary magnetosphere on the orbital distance.
The cosmic ray model (\textit{inside} the magnetosphere) is explained in Section \ref{sec-model}.
In Section \ref{sec-results}, the resulting flux of galactic
cosmic rays to the atmospheres of different extrasolar planets is discussed. 
We study the effect of orbital
distance via the stellar wind parameters (Section \ref{sec-distance}), tidal locking (Section \ref{sec-tidallocking}), 
planetary mass and planetary composition (Section \ref{sec-type}). 
Potential implications are briefly discussed in Section \ref{sec-implications}. 
Section \ref{sec-conclusions} closes with some concluding remarks.

\section{The planetary situation} \label{sec-planet}

\subsection{Tidal locking} \label{sec-tidallocking}

Tidal locking (i.e.~the synchronisation of the planetary rotation period
to the planetary orbital period)
occurs for planets orbiting very close to their parent star. 
The relevance of this process for terrestrial extrasolar planets 
is discussed in this section. This is an update to the corresponding
section of \citet{Griessmeier05a}.

A planet with angular velocity $\omega_\text{i}$ at $t=0$ (i.e.~after formation) will gradually lose
angular momentum, until the angular velocity reaches a constant value 
$\omega_\text{f}$ at $t=\tau_{\text{sync}}$.
For a planet with a mass of $M\p$ and radius of $R\p$ around a star of mass 
$M\sstar$, $\tau_{\text{sync}}$ can be expressed as 
\citep[e.g.][Appendix B]{Murray99,Griessmeier07AA}:
\begin{equation}
        \tau_{\text{sync}} \approx \frac{4}{9} \alpha \, Q\p' \left( \frac{R\p^3}{GM\p} \right)
        \left( \omega_\text{i}-\omega_\text{f} \right)
        \left( \frac{M\p}{M\sstar} \right)^2
        \left( \frac{d}{R\p} \right)^6
	\label{eq:locking}
\end{equation}
The parameters $\alpha$, $Q\p'$, $\omega_{i}$ and 
$\omega_{f}$ required to calculate the timescale for tidal locking of a
terrestrial planet are presented further below.

Eq.~(\ref{eq:locking}) shows that the timescale for tidal locking 
strongly depends on the distance ($\tau_{\text{sync}}\propto d^{6}$). Thus, a 
planet in a close-in orbit around its central star is subject to strong 
tidal interaction, leading to gravitational locking on a very short timescale. 
Other important factors for the tidal locking timescale 
are the stellar mass and the planetary structure.

\subsubsection{Structure parameter $\alpha$}

The constant $\alpha$ depends on the internal mass distribution within the
planet. 
It is 
defined by $\alpha=I/(M\p R\p^2)$,  
where $I$ is the planetary moment of inertia. 
For a sphere of homogeneous density, $\alpha$ is equal to 2/5. For planets, generally  
$\alpha\le 2/5$.
For the Earth, the 
structure parameter $\alpha$ is given by $\alpha=1/3$ 
\citep{Goldreich66}. In the following, this value will be used.

\subsubsection{Tidal dissipation factor $Q_\mathrm{p}'$} \label{sec:Q':terr}

$Q\p'$ is the modified $Q$-value of the planet. It is defined by
\citep{Murray99}
\begin{equation}
        Q\p'=\frac{3Q\p}{2k_{2,\text{p}}}, \label{eq:k_2}
\end{equation}
where $k_{2,\text{p}}$ is the Love number of the planet. $Q\p$ is the planetary
tidal dissipation factor (the larger it is, the smaller is the tidal 
dissipation), defined by \citet{MacDonald64} and \citet{Goldreich66}.

We use $k_{2,\text{p}}=0.3$ for the Earth 
\citep{MacDonald64,Murray99}. This value will also be used for the ``small
super-Earth'' and the ``Ocean Planet'' case. 
For the Earth, a value of $Q\p\approx12$ can be determined from the
measured secular acceleration of the moon \citep{Goldreich66,Murray99}.
This value is relatively small when compared to other terrestrial planets, where $Q\p$ is 
typically of the order of $10^2$.
This is probably due to the fact that, for the Earth, much energy is dissipated in the shallow 
seas \citep{MacDonald64,Hubbard84,Kasting93,Murray99}. 
In the past, when the continents were joined, the value of $Q\p$ was probably larger
\citep{Peale99,Murray99}.
For the case of the ``small super-Earth'' and the ``Ocean Planet'' (see below), 
$Q\p$ is set to 
the value usually assumed for Venus and Mercury
\citep[$Q\p\approx100$,][]{Murray99}. 

From $k_{2,\text{p}}$ and $Q\p$, the required value of $Q\p'$ can be obtained 
using eq.~(\ref{eq:k_2}). We find $Q\p'=60$ for Earth and $Q\p'=500$ for both
the ``small super-Earth'' and ``Ocean Planet'' cases.

\subsubsection{Initial rotation rate $\omega_\mathrm{i}$}

The initial rotation rate $\omega_\text{i}$ of a terrestrial planet is a poorly known quantity 
\citep[see, e.g.][]{Correia03a}. It certainly depends on the details of the planetary 
formation and can be strongly influenced by processes like migration or impacts.
Therefore, two limits for $\omega_\text{i}$ are considered: 
\begin{itemize}
	\item 	A relatively high initial rotation rate as suggested for the 
		early Earth-Moon system, with $\omega_\text{i}=1.83\, \omega\E$ 
		(with the current rotation period of the Earth
		$\omega\E=7.27\cdot10^{-5} $ s$^{-1}$) corresponding
		to a length of day of 13.1 h \citep{MacDonald64}
	\item 	A lower rotation rate with $\omega_\text{i}=0.80\, \omega\E$
		corresponding to a day of 30 h.
\end{itemize}
Note that a primordial rotation period of the order of 10 h is consistent with the relation 
between the planetary angular momentum density and planetary mass observed in the 
solar system \citep[][Chapter~4]{Hubbard84}.

\subsubsection{Final rotation rate $\omega_\mathrm{f}$}

As far as eq.~(\ref{eq:locking}) is concerned, the final rotation rate 
$\omega_\mathrm{f}$ can be neglected, at least
 for the planets of interest in this work
\citep{GriessmeierPHD06}. 

\subsubsection{Tidal locking: Results for exoplanets}

Eq.~(\ref{eq:locking}) allows us to calculate the timescale for tidal
locking for terrestrial exoplanets.
Subsequently, the planets can be classified as either ``tidally locked'', 
``potentially tidally locked'' or ``unlocked''. 
The upper and lower boundaries of the potentially locked region are determined
by the conditions $\ts=100$ Myr and $\ts=10$ Gyr, respectively. To account for
the uncertainty of the initial rotation rate $\omega_\text{i}$, the lower
boundary is calculated with $\omega_\text{i}=1.83\,\omega\E$ (i.e.~a rotation
period of 13.1 h), and the upper boundary is calculated with
$\omega_\text{i}=0.80\,\omega\E$  (corresponding to a rotation period of 30 h). 
This increases the area of the ``potentially locked'' region.

\begin{figure}[ht]
  \begin{center}
\centerline{\includegraphics[width=1.0\linewidth]
{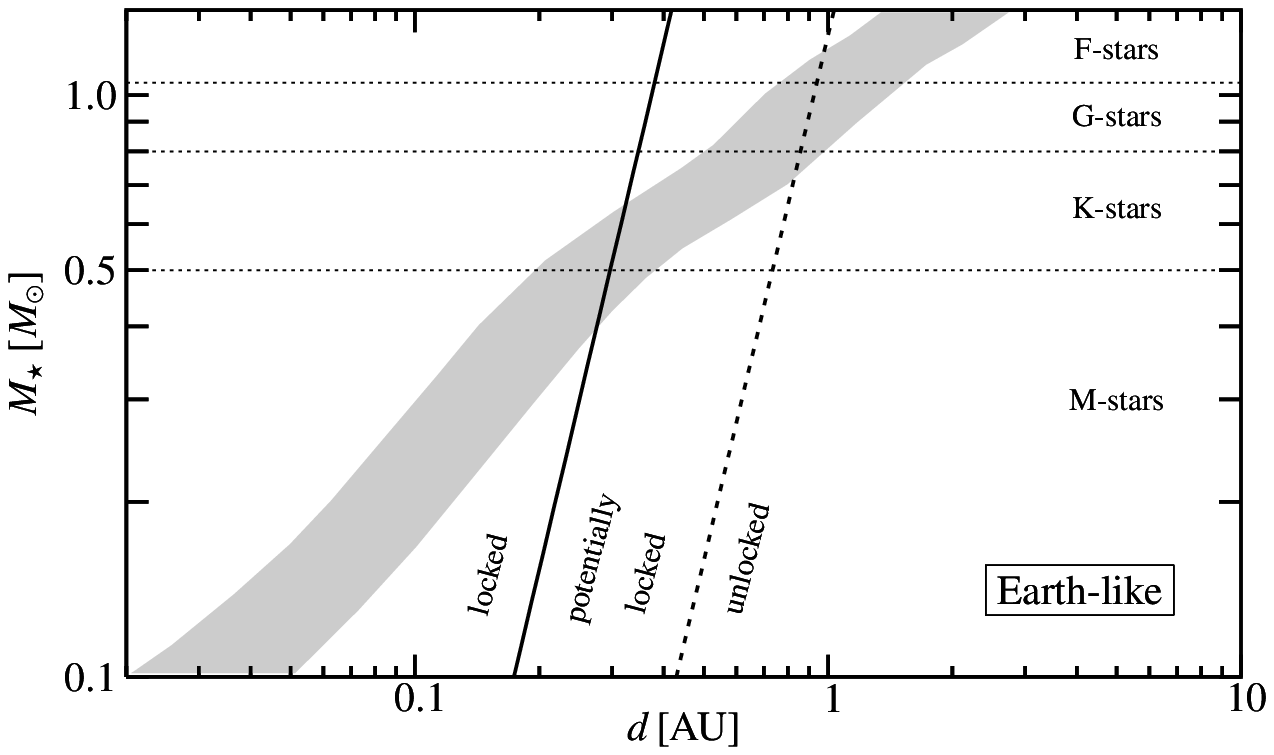}}    
\caption{
Tidally locked (left) vs. freely rotating (right) regime for Earth-like planets as a function 
of orbital  distance $d$ and mass $M\sstar$ of the host star. 
The shaded area gives the location of the continuously habitable 
zone (CHZ). Updated from \citet[][]{Griessmeier05a}.
\label{fig:tidallocking}}
\end{center}  
\end{figure}

In Figure \ref{fig:tidallocking}
\citep[updated from][]{Griessmeier05a}, 
the grey-shaded area gives the location of the 
continuously habitable zone (CHZ) according to \citet{Kasting93}. 
The lines delimit different parameter regimes for an Earth-like planet:
All planets to the 
right of the dashed line 
are considered as tidally locked
while 
planets to the left of the solid line  
are supposed to be freely rotating.
One can see that all
Earth-like planets inside the CHZ of M stars can be considered to be 
tidally locked. 

\begin{figure}[tb] 
\begin{center}
     \includegraphics[width=0.95\linewidth]
     {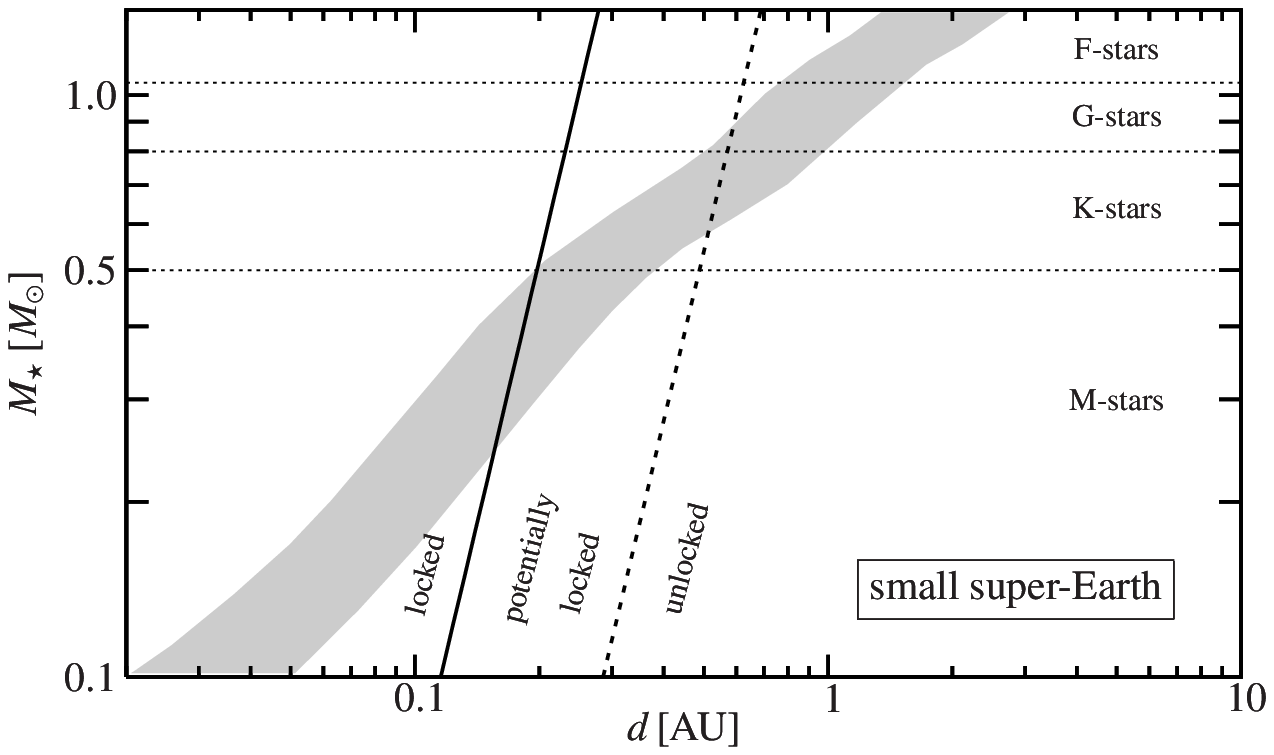}
\caption{
Tidally locked (left) vs. freely rotating (right) regime for a ``small
super-Earth''
as a function 
of orbital  distance $d$ and mass $M\sstar$ of the host star. 
The shaded area gives the location of the continuously habitable 
zone (CHZ).
\label{fig:tidallocking:LargeEarth}}
\end{center}
\end{figure}

\begin{figure}[tb] \begin{center}
     \includegraphics[width=0.95\linewidth]
     {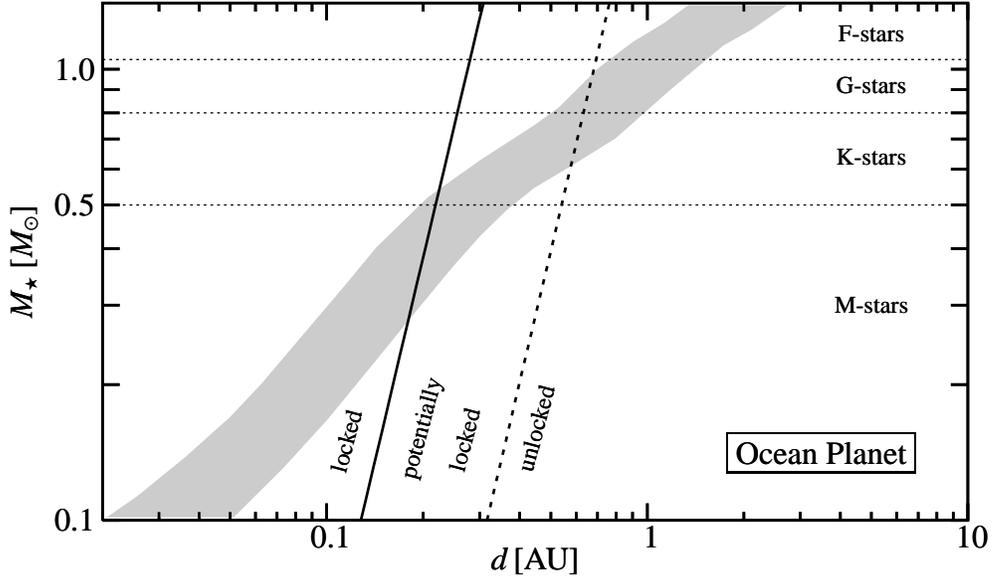}
\caption{Tidally locked (left) vs. freely rotating (right) regime for an ``Ocean Planet''
as a function 
of orbital  distance $d$ and mass $M\sstar$ of the host star. 
The shaded area gives the location of the continuously habitable 
zone (CHZ).
\label{fig:tidallocking:OceanPlanet}}
\end{center}
\end{figure}

In addition to the current Earth, two additional configurations are examined:
\begin{itemize}
	\item	The ``small
		super-Earth'' case (described in more detail in Section 
		\ref{sec-results})
		is depicted in Fig.~\ref{fig:tidallocking:LargeEarth}. It can be
		seen that the larger value of $Q\p'$ slightly decreases the 
		``tidally locked'' regime.
	\item	The ``Ocean Planet'' case (described in more detail in Section
		\ref{sec-results}) is depicted in 
		Fig.~\ref{fig:tidallocking:OceanPlanet}, with similar results
		than for the ``small
		super-Earth'' case.
\end{itemize}

As a typical test case, a terrestrial exoplanet at an orbital distance of 0.2 AU around a star
of 0.5 $M\sun$ will be studied.
Figs.~\ref{fig:tidallocking} to \ref{fig:tidallocking:OceanPlanet}
clearly show that such a planet, regardless of its precise mass, radius, and composition, is
very likely to be tidally locked. 
The reduced rotation rate of tidally locked planets is supposed to have 
important consequences for the planetary magnetic dipole moment, which is 
discussed in Section \ref{sec-magmoment}.

\subsection{Planetary magnetic moment} \label{sec-magmoment}

The planetary magnetic dipole moment can be estimated from different scaling 
laws \citep{Griessmeier05a}.
It can be shown that for the slowly rotating tidally locked planets, 
the magnetic moment is much 
smaller than for freely rotating planets. For an Earth-like planet in an orbit 
of 0.2 AU around a star with 0.5 stellar masses, the magnetic moment is
expected to lie in the range 
$0.02 \mathcal{M}\E<\mathcal{M}<0.15 \mathcal{M}\E$, where $\mathcal{M}\E$ is 
the value of Earth's current magnetic moment \citep{Griessmeier05a}. 
In the following, the maximum value of $0.15 \mathcal{M}\E$ will be adopted to 
obtain a
lower limit for the cosmic ray flux to the atmosphere. 
Magnetic moments resulting for a few other configurations are given in Table 
\ref{tab:Rs:cosmicrays}.

\subsection{Stellar wind}\label{sec-stellarwind}

It is known that at the close orbital distances of M-star habitable zones the stellar wind has 
not yet reached the quasi-asymptotic velocity regime.  
Because of the low stellar wind velocity, the planetary magnetosphere is less
strongly compressed at such distances than one would expect from stellar wind
models with a constant velocity.
To capture this behaviour realistically 
and to correctly describe the flux of galactic cosmic rays into the atmospheres of close-in exoplanets, we
require a model with a radially dependent stellar wind velocity.

For slowly rotating stars, the stellar wind may be described by the solution of the purely 
hydrodynamic, isothermal model of \citet{Parker58}. 
In this model, the interplay between stellar gravitation and pressure gradients leads to 
a supersonic flow for sufficiently large substellar distances $d$ 
\citep[see e.g.][]{Parker58, Mann99, Proelss04}. 
In accordance with the observations, this 
model describes a solar wind with low velocity and large acceleration near the sun, whereas at 
larger distances the velocity is large and the acceleration strongly decreases.

The procedure 
to obtain the stellar wind velocity 
$v\eff(d,M_\star,R_\star)$ and density $n(d,M_\star,R_\star)$ at the location of an 
exoplanet (i.e.~at distance
$d$) for a host star of given mass $M_\star$, and radius $R_\star$ 
consists of the following:
\begin{enumerate} 
	\item[(1)] 
		The stellar mass loss rate is assumed to be proportional to its surface area:
		\begin{equation}
			\dot{M}_\star=\dot{M}_\odot \frac{R_\star^2}{R_\odot^2},
			\label{eq:dotMstar}
		\end{equation}
		where the subscripts $_\odot$ and $_\star$ denote solar and stellar properties, respectively. 
    	\item[(2)] A Parker-like stellar wind model 
		is used to find $n(d)$ and $v(d)$ as a function of the distance to the star.
          	The coronal 
		temperature $T_{\text{corona}}$ is adjusted until the stellar wind velocity at 
		1 AU corresponds to the value which was obtained in step 1. 
		With this value of $T_{\text{corona}}$, $v$ can be determined for any value of
		$d$ by solving Parker's wind equation.
		Thus, $v(d)$ is obtained.
        \item[(3)] Because of $\dot{M}_\star=4\pi d^2 \,n(d)v(d)m$ (where $m$ is the proton 
		mass), 
		the density $n(d)$ is obtained by dividing the stellar mass loss
		$\dot{M}_\star$ obtained in step 1 by $4\pi d^2 \,v(d)m$, where $v(d)$ was 
		obtained in step 2.
	\item[(4)] The effective stellar wind velocity relative to the planet $v\eff$ is calculated from 
		$v$ (as obtained in step 2 above) and the planetary orbital velocity 
		($v\eff=\sqrt{v^2+v^2_\text{orbit}}$).	
\end{enumerate}
This procedure is similar to that of \citet{Griessmeier05a}, with the difference that here we keep the stellar age fixed (at 4.6 Gyr) to compare stars of identical ages.

\subsection{Magnetospheric model}\label{sec-magnetosphere}

The magnetosphere is modelled as a cylinder topped by a half-sphere 
\citep{Voigt81,Stadelmann04,Griessmeier05a}. 
A closed magnetosphere is assumed, i.e.~field lines are not
 allowed to pass
through the magnetopause.
Within this model, the magnetic field is defined for any point inside 
the magnetosphere as soon as the planetary magnetic moment and the size of the
magnetosphere are prescribed.
The size of the
magnetosphere is characterised by the magnetopause standoff distance $R_s$, i.e. the
extent of the magnetosphere along the line connecting the star and the planet.
$R_s$ can be obtained from the pressure equilibrium at the
substellar point. This pressure balance includes the stellar wind ram pressure, the stellar wind
thermal pressure of electrons and protons, and the planetary magnetic field pressure:
\begin{equation}
	m n v\eff^2+2 \, nk_BT=
	\frac{\mu_0f_0^2\mathcal{M}^2}{8\pi^2 R_s^6}. \label{eq:pressureequilibrium}
\end{equation}
Here, $f_0=1.16$ is the form factor of the magnetosphere and includes the magnetic field caused
by the currents flowing on the magnetopause \citep{Voigt95,Griessmeier04}. 
$\mathcal{M}$ is the planetary magnetic dipole moment as discussed in section 
\ref{sec-magmoment}. 

For a given planetary orbital distance $d$, only the magnetospheric magnetic pressure is a 
function of the distance to the planet, while the other contributions are constant. 
Thus, from the pressure equilibrium eq.~(\ref{eq:pressureequilibrium}) the standoff distance 
$R_s$ is found to be:
\begin{equation}
	 R_s = 
	 \left[ \frac{\mu_0f_0^2\mathcal{M}^2}
	 {8\pi^2 \left(m n v\eff^2+2 \, nk_BT\right)} \right]^{1/6}.
	 \label{eq:Rs}
\end{equation}
Table \ref{tab:Rs:cosmicrays} lists the standoff distances $R\sub$ (in planetary radii $R_p$) for different planetary 
configurations as well as the magnetic field strength at the location of the magnetopause, $B(R\sub)$.

\subsection{Cosmic ray flux outside the magnetosphere as a function of orbital distance}
\label{sec-distance1}

In the solar system, the cosmic ray flux is known to increase with increasing heliocentric 
distance. This is
attributed to effects such as diffusion, convection, adiabatic deceleration as well as gradient and
curvature drifts
\citep[see, e.g.][]{Kallenrode00, Heber06}. The
theoretical basis for this approach is described by \citet[][section 2.3]{Fichtner01}.
In this section, we study how strong such effects are in the inner heliosphere to
see how much smaller the cosmic ray flux outside the magnetosphere of a 
close-in exoplanet (e.g.~at 0.2 AU) may be 
compared to one in an Earth-like orbit. The penetration of these particles \textit{through}
the planetary magnetosphere is described in section \ref{sec-results}. 

The radial gradient of the cosmic ray intensity $J$ depends on 
the heliocentric distance $d$. Between 1 and 42 AU, \citet{McDonald92}
approximate it by:
\begin{equation}
	\frac{1}{J}\frac{dJ}{dd}=G_0 \left( \frac{d_e}{d}\right)^\alpha.
	\label{eq:cr-radial}
\end{equation}
During solar minimum, $\alpha\approx0.7$  
for cosmic ray protons (the exact value depends on the particle energy), and $G_0\approx 0.2
\text{ AU}^{-1}$. 
The orbital distance of the Earth is denoted by $d_e= 1$ AU.

\begin{figure}[t] \begin{center}
     \includegraphics[width=\linewidth]
	{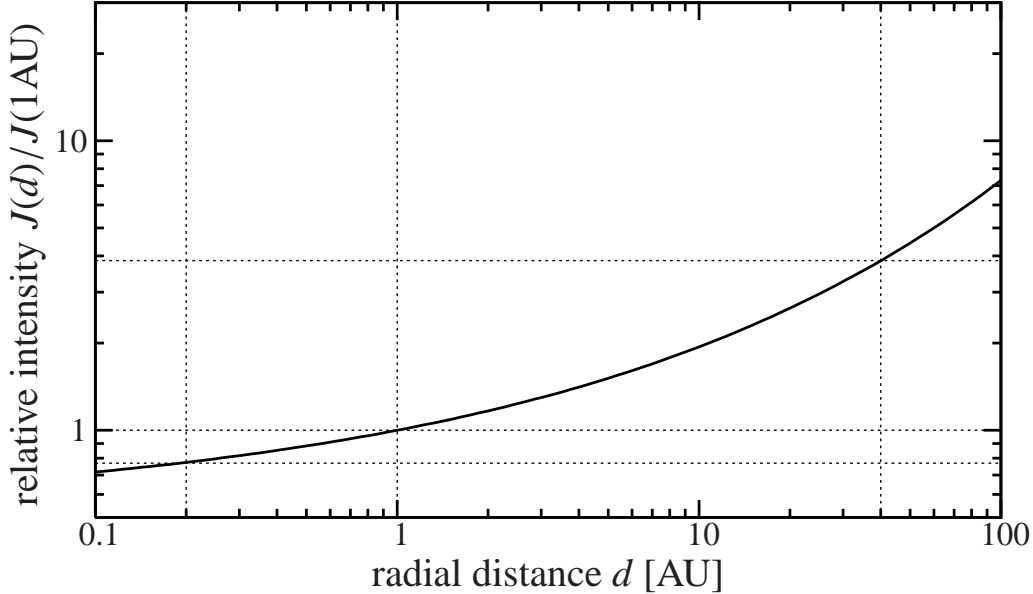}
     \caption{Relative cosmic ray flux in the heliosphere as a 
     function of heliocentric distance.
    \label{fig:cr-gradient}} 
\end{center}
\end{figure}

Eq.~(\ref{eq:cr-radial}) is equivalent to a relative cosmic ray flux of:
\begin{equation}
	J_{\text{rel}}(d)=J_0 \exp \left[  G_0 \left( d^{1-\alpha}\right) \frac{d_e^\alpha}{1-\alpha} \right],
\end{equation}
where $J_0$ is determined by the condition $J_{\text{rel}}(d_e)=1$.
Figure \ref{fig:cr-gradient} shows $J_{\text{rel}}(d)$.

With this, we can compare the cosmic ray flux at different distances.
We find that $J(0.2\text{AU})\approx0.8 \cdot J(1.0\text{AU})$, and 
$J(40\text{AU})\approx4 \cdot J(1.0\text{AU})$. For distances below 1.0 AU, the particle flux varies only 
weakly, so that we can neglect this effect when calculating fluxes of cosmic rays into the
atmospheres of close-in planets. 

Note that the radial gradient could be different for other stars than in the solar system. For a low mass
M star with a lower stellar wind density, we would expect the radial gradient of the cosmic ray
intensity to be smaller,
whereas it is likely to be larger for a star with a denser and faster stellar wind 
\citep[as would be
the case, e.g.~for a younger star, see][]{Wood02,Wood05,Griessmeier05a, Holzwarth07}.
While a quantitative computation of the cosmic ray gradients around other stars 
would be interesting, this is outside the scope of this paper. This question 
will be addressed in the future. 
For the moment, we will assume that the cosmic ray
modulation around other stars is not much larger than that found in the solar
system, and we will correspondingly neglected it in the remainder of this work.

\section{Cosmic ray protection}
\label{sec-model}

In order to quantify the protection of extrasolar Earth-like planets against galactic cosmic
rays, the motion of galactic cosmic protons through planetary magnetospheres is investigated numerically.

\subsection{Cosmic ray calculation}

In order to determine the impact of GCR protons in the energy range
64 MeV $< E <$ 8 GeV on the planetary atmosphere,
particle trajectories in the magnetosphere are analysed. 
Because no solution in closed form exists, this is only possible through the numerical
integration of many individual trajectories \citep{Smart00}.
For each particle energy (64~MeV, 128~MeV, 256 MeV, 512 MeV, 1024 MeV, 2048 MeV, 4096~MeV and 
8192~MeV) and for each magnetospheric configuration, over 7 million trajectories are calculated,
which correspond to protons with different starting positions and starting velocity directions.
The particles are launched from the surface of a sphere centred on the 
planet.
The radius of this sphere satisfies the condition $r\ge R\sub$, 
so that the particles 
are launched outside the magnetosphere (except for those arriving from the tailward direction).

Once the particles enter the magnetosphere, their motion is 
influenced by the planetary magnetic field.
This magnetic field is calculated from the 
magnetospheric model of Section \ref{sec-magnetosphere}.
Note that Table \ref{tab:Rs:cosmicrays} gives the \textit{maximum} values for the planetary magnetic 
moments.
For this reason, the results represent the lower limit for the cosmic ray flux
to the atmosphere. For a smaller magnetic moment, a larger cosmic ray flux is possible.
The trajectories
are calculated using the numerical leapfrog method.
For each energy, all particles are counted which reach the atmosphere (described by a
spherical shell one hundred kilometres above the planetary surface,
i.e.~$R_\text{a}=R\p+100$ km).
The impact of particles on the planetary atmosphere is quantified by the
energy spectrum, which is defined in Section \ref{sec:energyspectrum}.
More details on the numerical calculation of the cosmic rays 
trajectories can be found in \citet{Stadelmann04}.

Note that for the planetary system we implicitly assume an environment similar
to that of the solar neighbourhood. Planets in other
regions of the galaxy
(e.g.~those encountering a different interstellar medium, or those
close to the galactic centre) may  be subject to very different cosmic ray 
fluxes \citep[e.g.][]{Scherer08}.

\subsection{Cosmic ray energy spectrum}
\label{sec:energyspectrum}

\begin{figure}[ht]
  \begin{center} 
\includegraphics[width=\linewidth]
{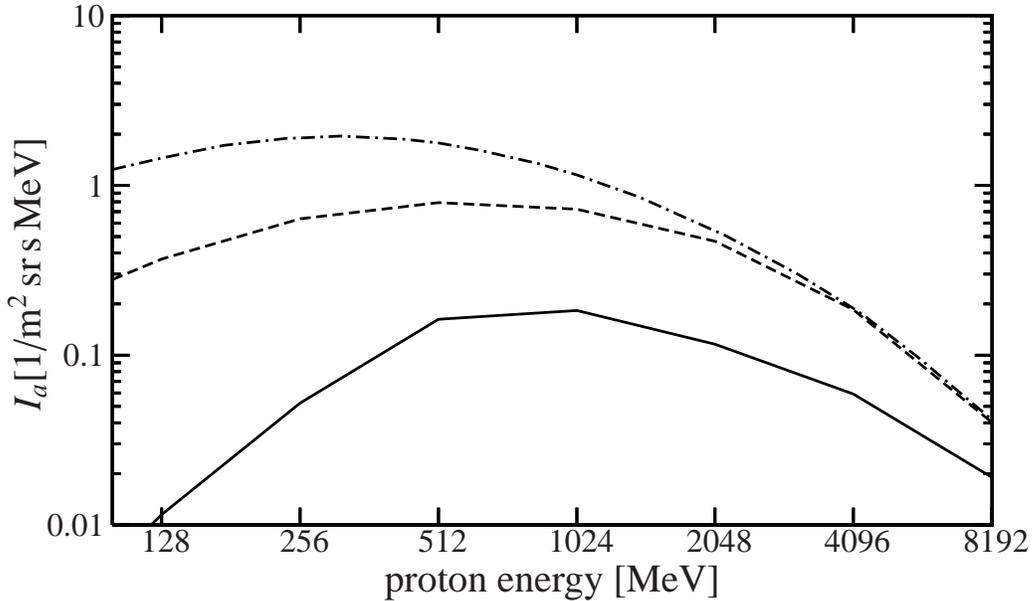}
\caption{
Dash-dotted line: energy spectrum outside the magnetosphere, according to
\citet{Seo94}.
Dashed line: energy spectrum of cosmic ray protons impacting the atmosphere 
(100 km above the surface) of an Earth-like planet at 0.2 AU around a 0.5 $M_\odot$ K/M star. 
Solid line: energy spectrum of cosmic rays impacting the atmosphere of
the Earth.
From \citet[][]{Griessmeier05a}.
\label{fig:energyspectrum}}
\end{center}
\end{figure}

The cosmic ray \textit{energy spectrum} is determined in the following way: for a given particle
energy, the fraction of 
particles reaching the planetary atmosphere is registered. 
This value is compared to the fraction of
particles reaching the atmosphere of an 
identical, but unmagnetised planet.
The resulting magnetospheric filter function is multiplied with the cosmic ray energy 
spectrum outside the magnetosphere, which was taken from \citet{Seo94}. 
Fig.~\ref{fig:energyspectrum} shows the reference energy spectrum of \citet{Seo94} as a
dash-dotted line.
The energy spectrum on top of the Earth's atmosphere is shown as a solid line, whereas the dashed
line corresponds to the cosmic ray energy spectrum for an Earth-like exoplanet in an orbit of 
0.2 AU around a 0.5 $M_\odot$ K/M star.

\section{Comparison of GCR fluxes for eight scenarios}
\label{sec-results}

In the following, the flux of cosmic rays at the top of the atmosphere
will be compared for eight different planetary configurations. 
The 
energy spectrum is determined for the following cases: 
\begin{itemize}
	\item Case 1: ``Unmagnetised case''. 
	      This is the (reference) cosmic ray energy spectrum outside any
	      planetary magnetosphere \citep[interplanetary case, taken
	      from][]{Seo94}. 
	      At the same time, this corresponds to the flux arriving at
	      the top of the atmosphere of a totally
	      unmagnetised planet (i.e.~without even induced magnetic fields). 
	      This case is valid for any stellar mass.
	      In the following figures, it is represented by a dash-dotted line. 
	\item Case 2: ``(Unlocked) Earth case''. The case of the magnetosphere
	      of today's Earth (i.e.~at 1.0 AU of a sun-like star
	      with $M\sstar=1.0 M\sun$).  
	      The Earth's magnetic field is represented by a zonal dipole with a
	      Gaussian coefficient of $g_{10}=31$ $\mu$T. 
	      In all figures, 
	      this case is represented by a solid line.
	\item Case 3: ``Locked exoplanet case''. The case of the magnetosphere of an extrasolar planet 
	      in an orbit of 
	      0.2 AU around an K/M star of 0.5 solar masses. Due to tidal locking, the magnetic 
	      moment is reduced to 15 \% of the Earth's magnetic moment
	      (see Section \ref{sec-magmoment}). 
	      For this situation, a dashed line is used.
	\item Case 4: (Hypothetical) ``unlocked exoplanet case''. Identical to 
	      the ``locked exoplanet case'' (e.g.~located at 0.2 AU
	      around a sun-like star with $M\sstar=0.5 M\sun$), with the only
	      difference being the magnetic moment: Here we assume
	      that the planet is 
	      \textit{not} tidally locked and has a strong magnetic moment ($\MM=1.0 \MM\E$). 
	      Note that this case is not realistic
	      (unless the planet was brought to this position only recently), but it is
	      instructive to compare the influence of the stellar proximity and the small
	      planetary magnetic moment. In the figures, this case will be shown by filled triangles.
	\item Case 5: (Hypothetical) ``locked Earth case''. Identical to the 
	      ``(unlocked) Earth case'' (e.g.~~at 1.0 AU distance from
	      a sun-like star with $M\sstar=1.0 M\sun$),
	      but assuming that the planet only has a small magnetic moment
	      ($\MM=0.15 \MM\E$). 
	      This case 
	      can also be seen as a ``weakly magnetised Earth case''.
	      Similarly to the ``unlocked exoplanet case'', this case will be used
	      to disentangle the effect of stellar proximity and of small magnetic moment. 
	      For this case, circles are used.
	\item Case 6: ``Small super-Earth case''
	      \citep[6 Earth masses, 1.63 Earth radii, as modelled by][]{Leger04}.
	      Similar to the ``locked exoplanet 
	      case'' (case 3), but for a large terrestrial exoplanet, assuming tidal locking. 
	      As was mentioned before, the first planets of such size have now been detected. 
	      This case is represented by a double-dashed line.
	\item Case 7: ``Large super-Earth case'' 
	      \citep[10 Earth masses, 1.86 Earth radii, making use of the
	      scaling relations of][]{Valencia06}.
	      Similar to the ``locked exoplanet 
	      case'' (case 3) or the ``Small super-Earth case'' (case 6), but for a larger planetary mass, 				assuming tidal locking. This
	      planet is denoted by a triple-dashed line. Together with the ``locked exoplanet 
	      case'' and the ``small super-Earth case'', this will be used to analyse the
	      dependence of cosmic ray protection on the planetary mass.
	\item Case 8: ``Ocean planet case'' \citep[6 Earth masses, 2.0 Earth radii, 
	      as modelled by][]{Leger04}. Similar to the  ``Small super-Earth case'' (case 6), 
	     but for an Ocean Planet
              \citep[i.e. a planet with a significant fraction of water ice, see e.g.][]{Kuchner03,Leger04,Selsis07}. 
              For this situation, a dotted line is used.
\end{itemize}

For each of the above cases, the input parameters (the orbital distance
$d$ and the stellar mass $M\sstar$) and the resulting values (the magnetic 
moment $\mathcal{M}$, the standoff distance $R\sub$ and the magnetic field strength at the 
magnetopause $B(R\sub)$)
are given 
in Table \ref{tab:Rs:cosmicrays}.
The stellar mass has two functions: It determines the rotation rate of
tidally locked planets (case 3 and cases 5-8), and it defines the stellar
wind flux onto the magnetosphere, thus defining the size of the magnetosphere
via Eq. (3). For case 1 (no magnetosphere), the stellar mass does not matter.
As noted before, the magnetic moments (and thus also the standoff 
distances) are assumed to have the maximum allowed value. Thus the values obtained in 
this section represent a lower limit for the flux of cosmic ray protons to the planetary 
atmosphere. Table \ref{tab:Rs:cosmicrays} also includes the magnetic field strength at the 
magnetopause. 
By definition of the magnetic moment, $B(R\sub)\propto \mathcal{M}R\sub^{-3}$.
When the contribution of the thermal pressure in
eq.~(\ref{eq:pressureequilibrium})  is negligible, this results in  $B(R\sub)
\propto \sqrt{nv\eff^2}$, so that the magnetic field strength at the
magnetopause is identical e.g~for cases 3 and 4, but different for case 
5.  

\begin{table}[!h]
\begin{center} 
   \begin{tabular}{|l||c|c||c|c|c|}\hline
        Planet 
		& $d$ & $M\sstar$ & $\mathcal{M}$ & $R\sub$ & $B(R\sub)$
		\\
	& [AU] & [$M\sun$] & [$\MM\E$] & [$R\p$] & [nT]
		\\[4pt] \hline \hline
	Case 1: Unmagnetised                
		& any & any & $0$	& - & -
		\\[4pt] \hline	
	Case 2: (Unlocked) Earth                
		& 1.0 & 1.0 & $1.0$	& $9.91$ & $73$
		\\[4pt] \hline
	Case 3: Locked exoplanet                 
		& 0.2 & 0.5 & $0.15$ & $4.12$ & $150$
		\\[4pt] \hline
	Case 4: (Hyp.) unlocked exoplanet                 
		& 0.2 & 0.5 & $1.0$	& $7.81$ & $150$
		\\[4pt] \hline
	Case 5: (Hyp.) locked Earth            
		& 1.0 & 1.0 & $0.15$ & $5.23$ & $73$
		\\[4pt] \hline
	Case 6: Small super-Earth (6 $M\E$)           
		& 0.2 & 0.5 & $0.65$ & $4.15$ & $150$
		\\[4pt] \hline
	Case 7: Large super-Earth (10 $M\E$)            
		& 0.2 & 0.5 & $0.96$ & $4.13$ & $150$
		\\[4pt] \hline	
	Case 8: Ocean Planet            
		& 0.2 & 0.5 & $0.37$ & $2.80$ & $150$
		\\[4pt] \hline
   \end{tabular}
   \caption[Parameters for cosmic ray calculation]
   {Parameters for the different planetary configurations.
   $d$: orbital distance, $M\sstar$: stellar mass, 
   $\mathcal{M}$: planetary magnetic moment, $R\sub$: standoff distance, $B$: magnetic field at the magnetopause.  
   }
\label{tab:Rs:cosmicrays}
\end{center} 
\end{table}

The cosmic ray flux to the planetary atmosphere depends also on the stellar age (which
is a crucial parameter for the stellar wind velocity and density). This effect was discussed in 
\citet{Griessmeier05a}.

\subsection{Influence of orbital distance via the stellar wind parameters}
\label{sec-distance}

Previously, the cosmic ray flux to the atmosphere of a tidally locked Earth-like exoplanet around
a K/M type star with $M\sstar=0.5 M\sun$ has been calculated \citep{Griessmeier05a}. It was shown that such a 
planet (corresponding to our case 3) will be
subject to very high cosmic ray fluxes when compared to the Earth
(corresponding to our case 2). For particle energies below 200 MeV, the 
cosmic ray flux to the exoplanet was found to be up to one order of magnitude higher than on
Earth, and for energies above 2\nolinebreak\ GeV 
magnetospheric shielding is negligible for the exoplanet case. This can also be seen in 
Fig.~\ref{fig:energyspectrum}. 
As stated by \citet{Griessmeier05a}, two effects can contribute to
this high cosmic ray flux: (a) the reduced planetary magnetic dipole moment due to tidal locking 
(Section \ref{sec-magmoment}) 
and (b) the enhanced stellar wind ram pressure at small orbital distances (Section 
\ref{sec-stellarwind}). 
Both\footnote{In principle, a third effect plays a role,
namely the stellar mass and radius, which determine the stellar wind (e.g.~via 
Eq.~(\ref{eq:dotMstar})). However, if a solar mass star was used in case 3, the
planetary magnetosphere would be even more compressed, which does not decrease
the cosmic ray flux. Thus, the two other effects have to be responsible 
for the large difference in cosmic ray fluxes.} 
contribute to the magnetospheric compression, which in turn 
determines the flux of high energy cosmic ray particles onto the planetary atmosphere. Here,
we compare the relative importance of these two effects.

\begin{figure}[ht]
\begin{center} 
\includegraphics[width=\linewidth]
{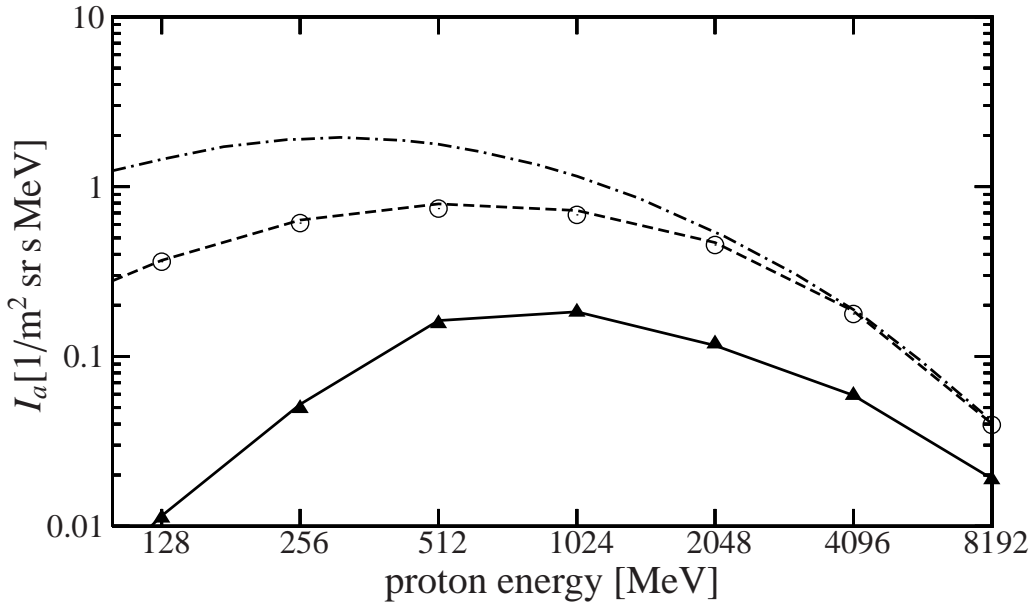}  
	\caption[Energy spectrum: influence of tidal locking]
	{Cosmic ray energy spectrum.
	Dash-dotted line: energy spectrum outside the magnetosphere (case 1, ``unmagnetised case'').
	Dashed line: ``locked exoplanet case'' (case 3). 
	Circles: hypothetical ``locked Earth case'' (case 5).
	Solid line: ``(unlocked) Earth case'' (case 2). 
	Triangles: hypothetical ``unlocked exoplanet case'' (case 4).
	\label{fig:energylocking}}
\end{center}
\end{figure}

We first estimate the importance of the orbital distance via the stellar
wind parameters. 
In Fig.~\ref{fig:energylocking},
we first compare the cosmic ray
energy spectra of the two strongly magnetised cases (with $\MM=1.0 \MM\E$):
The case of the Earth (case 2, solid line) and
the hypothetical case of a strongly magnetised exoplanet
(case 4, triangles).
It can be seen that the cosmic ray energy spectra are identical 
at $d=$1.0 AU (case 2, solid line) and at $d=$0.2 AU (case 4, triangles).
We also compare the two weakly magnetised cases (with $\MM=0.15 \MM\E$):
the weakly magnetised exoplanet case with $d=$0.2 AU 
(case 3, dashed line) and the hypothetical case of the weakly magnetised 
Earth at $d=$1.0 AU (case 5, circles).
Again, despite the large difference in orbital distance (0.2 and 1.0 AU), 
Fig.~\ref{fig:energylocking} shows that the cosmic ray energy spectrum is 
identical for these two cases.

The distance-dependent stellar wind ram pressure does not seem to have any measurable 
influence on the
cosmic ray energy spectrum. This may seem surprising, as it clearly has some influence on the 
size of the planetary magnetosphere, see Table \ref{tab:Rs:cosmicrays} (column 5). 
However, Table \ref{tab:Rs:cosmicrays}, 
column 6 shows that the magnetic field at the magnetopause is \textit{increased} when the orbital distance 
of a planet is decreased while keeping the magnetic moment constant. This directly results
from the pressure balance in eq.~(\ref{eq:pressureequilibrium}).
Thus, when the orbital distance is decreased while  
the magnetic moment is kept fixed, the smaller size of the magnetosphere is compensated by a 
larger magnetic field at the magnetopause $B(R\sub)$, which keeps the cosmic ray shielding at a constant
level. In other words, $R\sub$, the size of the planetary magnetosphere, is not a sufficient indicator for the quality of 
the magnetospheric shielding of a planetary atmosphere.

\subsection{Influence of tidal locking}

\label{sec-tidallocking}

If the difference of the cosmic ray energy spectra between a locked exoplanet and the Earth cannot
be explained by the different orbital distances, the discrepancy lies in
magnetic moment differences. 
This can be seen by comparing the cosmic ray energy
spectrum of the Earth case (case 2, solid line) to the cosmic ray flux of a
hypothetical Earth twin planet with a reduced magnetic moment 
(case 5, circles), 
with all other parameters kept constant ($d=1.0$ AU, $M\sstar=1.0 M\sun$).
Fig.~\ref{fig:energylocking} shows that the energy spectrum is much higher in
the  weakly magnetised case ($\MM=0.15 \MM\E$, case 5, circles)
than in the strongly magnetised case ($\MM=1.0 \MM\E$, case 2, solid line). 
We also compare the two exoplanet cases ($d=0.2$ AU, $M\sstar=0.5 M\sun$), 
namely the ``locked exoplanet case'' (case 3, dashed line) and the ``unlocked
exoplanet case'' (case 4, triangles).
Again, it can be seen that the cosmic ray energy spectrum has much higher values
for the weakly magnetised case than in the strongly magnetised case.
Obviously, the influence of tidal locking, hence the 
reduced magnetic moment (and not the stellar wind ram pressure) 
is the decisive factor for the increased influx of cosmic ray particles in case
3 (when compared to case 2). 
The similarity of cases 2 and 4 (and of cases 3 and 5) also shows that protons
of the chosen energy range mostly feel the planetary dipole field, and that the
field of the magnetpause currents has little influence \citep[eq. 20]{Vogt07}.

Similarly to a change in orbital distance, the planetary magnetic moment 
directly influences the size of the
planetary magnetosphere (see Table \ref{tab:Rs:cosmicrays}, column 5). When the magnetic moment is changed, 
however, the magnetic
field strength at the magnetopause (shown in Table \ref{tab:Rs:cosmicrays}, column 6) 
does not differ (compare cases 2 and 5, and cases 3 and 4, respectively).
Thus, when $B(R\sub)=$const,
the quality of the magnetospheric shielding of the planetary atmospheres 
depends on the size of the magnetosphere $R\sub$, i.e.~it
is weaker for those planets with a smaller magnetosphere.

We conclude that, of the two effects suggested by \citet{Griessmeier05a}, only one (namely the
effect of tidal locking) has a measurable influence on the cosmic ray flux. The other effect
(stellar wind density variation with orbital distance) appears to have negligible influence.

\subsection{Influence of planetary type}

\label{sec-type}

For planets with different size or composition than the Earth, the planetary magnetic dipole 
moment and the size of the magnetosphere are different from the Earth-like case. 
A few such cases are shown in Table \ref{tab:Rs:cosmicrays} and discussed in this section.
 
\begin{figure}[ht] 
\begin{center}
\includegraphics[width=\linewidth]
{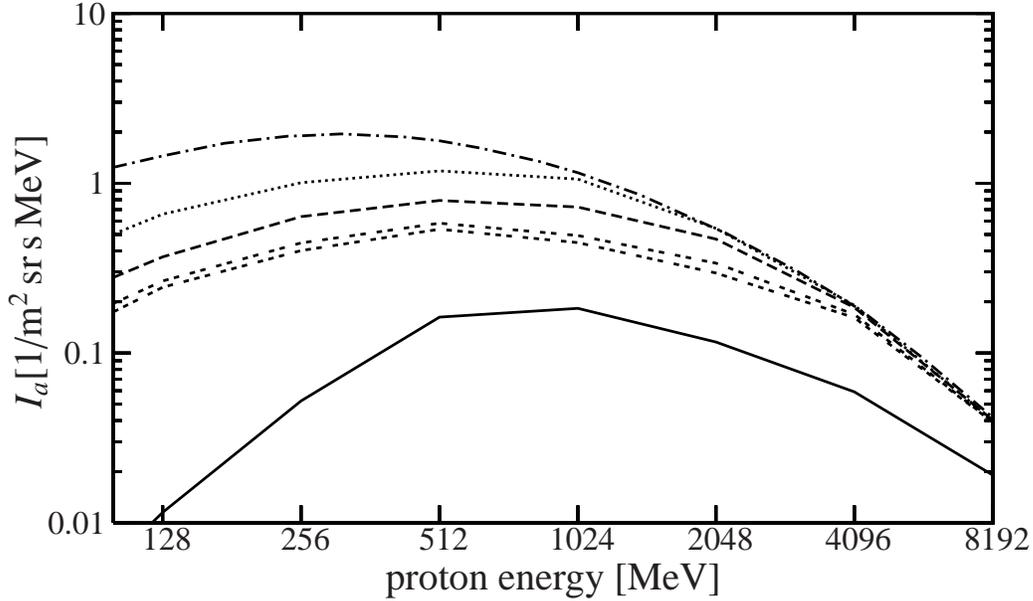}
     	\caption[Energy spectrum: influence of planetary type]
     	{Cosmic ray energy spectrum for different types of planets.
	Dash-dotted line: energy spectrum outside the magnetosphere (case 1, ``unmagnetised case'').
	Dotted line: ``Ocean Planet case'' (case 8).
	Dashed line: Earth-like planet (case 3, ``locked exoplanet case'').
	Double-dashed line: ``small super-Earth case'' (case 6).
	Triple-dashed line: ``large super-Earth case'' (case 7).
	Solid line: ``(unlocked) Earth case'' (case 2).
	\label{fig:energy46yLO}}
\end{center}
\end{figure}

Both ``super-Earth'' cases (cases 6 and 7) correspond to the case of a
terrestrial planet with a composition similar to that of the Earth, but with
higher planetary masses (6 and 10 Earth masses, respectively) and correspondingly larger 
radii \citep[1.63 and 1.86 Earth radii, see the models of][]{Leger04,Valencia06}.
For both cases, the expected magnetic moment
(Table \ref{tab:Rs:cosmicrays}, column 4) is larger than for an Earth-like
planet (case 3).
The larger magnetic moments lead to more extended magnetospheres (Table
\ref{tab:Rs:cosmicrays}, column 5).
Because the planetary radius is larger, however, the planetary atmosphere is
located at a larger distance from the planetary centre, 
so that the protection from 
galactic cosmic rays should not be expected to be much more efficient than for an Earth-like 
planet.
Figure \ref{fig:energy46yLO} shows that this is indeed the case: The energy spectra of the
``locked exoplanet case'' (case 3, dashed line), of the ``small super-Earth case'' (case 6, double-dashed line)
and of the ``large super-Earth case'' (case 7, triple-dashed line) are relatively similar. 
It can be seen that massive terrestrial planets do not necessarily have a much more efficient 
magnetospheric protection against galactic cosmic rays than an Earth-mass planet. 
 
The ``Ocean Planet case'' (case 8, dotted line) corresponds to a planet 
with a different composition: 17\% metals, 33\% silicates, and 50 \% water ice
\citep[this corresponds to the Ocean Planet modelled by][]{Leger04}. This
change in composition reduces the average density when compared to the
``super-Earth'' case of the same mass: The radius of a 6 Earth-mass Ocean Planet
is 2.0 Earth radii (as compared to the 1.63 Earth radii used above).
This moves the atmosphere further away from the centre of the
protecting dipole. 
Also, the reduced density leads to a somewhat reduced magnetic moment estimation.
This is reflected in the small standoff distance measured in planetary radii (see Table
\ref{tab:Rs:cosmicrays}) when compared to the ``small super-Earth case'' (case 6, double-dashed
line), and the cosmic ray flux through the magnetosphere can be expected to be larger.
These expectation is confirmed by the cosmic ray energy spectrum in Figure \ref{fig:energy46yLO}.
More cosmic ray particles reach the atmosphere for the ``Ocean Planet case'' (case 8, dotted line) than 
that of the ``small super-Earth case'' (case 6, double-dashed lines). Note that both planets have the same
mass (6 $M\E$), so that the difference is caused only by the difference in composition.
For energies above 100 MeV, the cosmic ray protection of an Ocean Planet (case 8) is weaker by a factor 
2-3 than for a rocky planet of similar mass (case 6). We conclude that the planetary composition can be a
potentially important factor concerning cosmic ray protection. 

\section{Implications}\label{sec-implications}

The increased flux of cosmic ray protons of galactic origin to the upper atmospheres of M-star
planets has implications for the atmospheric composition and thus for the remote detection of 
biomarkers. It will also influence the flux of (secondary) cosmic ray particles reaching the 
ground, which can be expected to have biological implications.  

\subsection{Atmospheric implications}

In the near future, atmospheres of extrasolar planets will
be studied from the point of view of possible biological
activity. ``Atmospheric biomarkers'' are compounds present
in an atmosphere which imply the presence of life and
which cannot be explained by inorganic chemistry alone.
The simultaneous presence of significant amounts of
atmospheric reducing gases (e.g. methane) and oxidising gases
(e.g. oxygen) as on the Earth is also an indicator
of biological activity \citep{Lovelock65,Sagan93}.
Good biomarkers include oxygen (produced by photosynthesis),
ozone (mainly produced from oxygen) and nitrous oxide
(produced almost exclusively from bacteria).

For the study of biomarkers in planetary atmospheres, it is important
to know all inorganic effects which can modify the abundances of these
molecules.
Modelling studies \citep[e.g.][]{Schindler00,Selsis02}
are required to rule out false conclusions in cases where inorganic
chemistry can mimic the presence of life. \citet{Grenfell07PSS}
looked at the response of biomarker 
chemistry on varying planetary parameters (e.g. orbital distance,
stellar type). Here, we will discuss the response of biomarkers to cosmic rays.

The influence of a high flux of GCRs to the atmospheres of planets in the habitable zone (HZ) around M-stars
was studied by \citet{Grenfell07AB}. 
When cosmic rays travel through an Earth-like atmosphere, they have sufficient energy to break
the strong $\mathrm{N_2}$ molecule and the chemical products react with oxygen to form nitrogen oxides
($\mathrm{NO_x}$).  If UV levels are intense, e.g. corresponding to the upper mesosphere and above on the Earth, then the $\mathrm{NO_x}$ cannot survive. $\mathrm{NO_x}$ affects biomarkers differently depending on altitude.
In a tropospheric environment, ozone is enhanced by $\mathrm{NO_x}$ due to the smog mechanism \citep{HaagenSmit52}.
In a stratospheric environment, ozone is depleted by catalytic cycles \citep{Crutzen70}. Once ozone is affected,
the other biomarkers can change too, because ozone affects the temperature profile (hence e.g. mixing
processes, chemical sources and sinks) and UV levels (hence photolysis rates) of other chemical species.

Using the cosmic ray spectra of our cases 1, 2 and 3, it was shown that GCRs may enhance the production of 
$\mathrm{NO_x}$ by a factor of three, which may modify
the abundances of biomarker molecules (especially water and ozone) by up to 25\% 
\citep{Grenfell07AB}.

It thus seems that most biomarkers are not strongly influenced by galactic cosmic rays for 
Earth-like planets in close orbits around M-stars. The case is different for solar energetic particles,
which can strongly modify the abundance of ozone in the planetary atmosphere \citep{Grenfell08}.
Such changes in atmospheric biomarker concentrations have to be taken into account when 
searching for biosignatures in the spectra of Earth-like exoplanets \citep{Selsis04}
by future space missions like DARWIN 
\citep{Fridlund04} or SEE-COAST \citep{Schneider06}. 

\subsection{Biological implications}

While a quantitative treatment is not yet available, 
the increased flux of cosmic rays at the top of the planetary atmosphere has 
not only implications for atmospheric chemistry and biomarkers, but also for the flux of secondary radiation on the surface.
When galactic cosmic rays of sufficiently high energy reach the planetary atmosphere, they 
generate showers of secondary cosmic rays.
Of the different components of such showers,
slow neutrons have the strongest influence on biological systems.  
The propagation of high energy cosmic ray particles 
also depends on the composition and density of the planetary atmosphere.
For an Earth-like atmosphere, the minimum energy which a proton must have to initiate a nuclear 
interaction that may be detectable at sea level is approximately 450 MeV \citep{Reeves92,Shea00}.
Terrestrial planets with dense atmospheres like Venus (100 bar surface
pressure) would be better shielded by the planetary atmosphere, 
so that no secondary radiation can reach the surface. On the other hand, for planets with thin
atmospheres like Mars (6 mbar surface pressure), 
the surface would probably be totally sterilised, even for a relatively low cosmic ray flux.

Keeping the atmospheric pressure constant at $\sim$1 bar, 
the flux at energies $>450$ MeV is strongly increased for the planetary situations presented in Section \ref{sec-results} (see Figs.~\ref{fig:energylocking} and 
\ref{fig:energy46yLO}), and a large increase of  secondary cosmic rays can be expected at the 
planetary surface.

Biological effects, namely an increase in cell fusion indices for different cell-lines, were
found to be significantly correlated with the neutron count rate at the Earth's surface 
\citep{Belisheva05,Griessmeier05a}.
Similar, but much stronger and more diverse effects were observed during large solar particle 
events, where solar cosmic rays dominate over galactic cosmic rays.
The effect of cosmic rays on cells and on DNA are potentially hazardous for life.
On the other hand, changes at the genetic level are a necessary condition for 
biological evolution, so that cosmic rays may also play a role in radiation-induced evolutionary
events.  

As tidally locked Earth-like exoplanets inside the HZ of low-mass K /
high-mass M stars
are only weakly protected against high energetic cosmic
rays, they can be expected to experience a higher surface neutron flux and stronger biological
effects than Earth-like planets with a strong magnetic moment.
For this reason, it may be more difficult for life to develop on the surface of 
planets around such stars.
It was also shown (Section \ref{sec-type}) that larger planets (super-Earths) are not much better protected 
against GCRs than smaller planets.

\section{Conclusion}\label{sec-conclusions}

The atmospheres of extrasolar planets with orbits in the habitable zone of M
stars are exposed to high fluxes of galactic cosmic rays. We have studied the
effect of orbital distance via the stellar wind parameters, tidal locking, planetary mass and planetary
composition on the cosmic ray fluxes. 
In previous works, the weak shielding found for M star planets was attributed to a combination of 
the enhanced stellar wind ram pressure with the reduced planetary magnetic dipole moment \citep{Griessmeier05a,GriessmeierESLAB05}. 
Here, we have shown that the variable stellar wind density at different orbital distance 
ranging from 0.2 to 1.0 AU (i.e. the CHZ of star with $0.5 M\sun<M\sstar<1.0
M\sun$).
 has little 
influence on the GCR flux. 
The planetary magnetic moment is the decisive parameter: Tidally
locked planets with a small magnetic moment have much higher particle fluxes than freely rotating planets with
stronger magnetic moments. 

We also find that more massive planets are not necessarily much better
protected against galactic cosmic rays, but that the planetary composition can play an important
role.

Generally speaking, to determine the magnetospheric shielding of a 
planetary atmosphere, it is not sufficient to calculate $R\sub$, the size of 
the planetary magnetosphere, but one has also to take into account $B(R\sub)$,
the magnetic field at the magnetopause. 

The above factors have to be considered for missions studying biosignatures in the spectra of 
Earth-like exoplanets for two reasons:
The enhanced flux of galactic cosmic rays may naturally generate or
destroy molecules typically regarded as biomarkers \citep{Grenfell07AB}, which has to be taken
into account in the analysis and interpretation of such spectra.
Also, as the habitability of M star planets may be affected, 
future observation campaigns should ensure that 
enough non-M stars are in the sample to maximise the likelihood of a positive detection. 

\ack

This study was supported by the International Space Science
Institute (ISSI) and benefited from the ISSI Team 
``Evolution of Exoplanet Atmospheres and their Characterisation''.
J. L. Grenfell and H. Lammer acknowledge the Helmholtz-Gemeinschaft 
as this study was also carried out within the framework of the Helmholz 
research project ``Planetenentwicklung und Leben: Die Suche nach Leben im Sonnensystem 
und dar\"{u}ber hinaus''.
We would also like to thank the referees for their comments and suggestions which helped
to improve the paper.



\begin{thebibliography}{64}
\expandafter\ifx\csname natexlab\endcsname\relax\def\natexlab#1{#1}\fi
\expandafter\ifx\csname url\endcsname\relax
  \def\url#1{\texttt{#1}}\fi
\expandafter\ifx\csname urlprefix\endcsname\relax\def\urlprefix{URL }\fi

\bibitem[{{Beaulieu} et~al.(2006){Beaulieu}, {Bennett}, {Fouqu{\'e}},
  {Williams}, {Dominik}, {Jorgensen}, {Kubas}, {Cassan}, {Coutures},
  {Greenhill}, {Hill}, {Menzies}, {Sackett}, {Albrow}, {Brillant}, {Caldwell},
  {Calitz}, {Cook}, {Corrales}, {Desort}, {Dieters}, {Dominis}, {Donatowicz},
  {Hoffman}, {Kane}, {Marquette}, {Martin}, {Meintjes}, {Pollard}, {Sahu},
  {Vinter}, {Wambsganss}, {Woller}, {Horne}, {Steele}, {Bramich}, {Burgdorf},
  {Snodgrass}, {Bode}, {Udalski}, {Szyma{\'n}ski}, {Kubiak}, {Wi{\c e}ckowski},
  {Pietrzy{\'n}ski}, {Soszy{\'n}ski}, {Szewczyk}, {Wyrzykowski},
  {Paczy{\'n}ski}, {Abe}, {Bond}, {Britton}, {Gilmore}, {Hearnshaw}, {Itow},
  {Kamiya}, {Kilmartin}, {Korpela}, {Masuda}, {Matsubara}, {Motomura},
  {Muraki}, {Nakamura}, {Okada}, {Ohnishi}, {Rattenbury}, {Sako}, {Sato},
  {Sasaki}, {Sekiguchi}, {Sullivan}, {Tristram}, {Yock}, and
  {Yoshioka}}]{Beaulieu06}
{Beaulieu}, J.-P., {Bennett}, D.~P., {Fouqu{\'e}}, P., {Williams}, A.,
  {Dominik}, M., {Jorgensen}, U.~G., {Kubas}, D., {Cassan}, A., {Coutures}, C.,
  {Greenhill}, J., {Hill}, K., {Menzies}, J., {Sackett}, P.~D., {Albrow}, M.,
  {Brillant}, S., {Caldwell}, J.~A.~R., {Calitz}, J.~J., {Cook}, K.~H.,
  {Corrales}, E., {Desort}, M., {Dieters}, S., {Dominis}, D., {Donatowicz}, J.,
  {Hoffman}, M., {Kane}, S., {Marquette}, J.-B., {Martin}, R., {Meintjes}, P.,
  {Pollard}, K., {Sahu}, K., {Vinter}, C., {Wambsganss}, J., {Woller}, K.,
  {Horne}, K., {Steele}, I., {Bramich}, D.~M., {Burgdorf}, M., {Snodgrass}, C.,
  {Bode}, M., {Udalski}, A., {Szyma{\'n}ski}, M.~K., {Kubiak}, M., {Wi{\c
  e}ckowski}, T., {Pietrzy{\'n}ski}, G., {Soszy{\'n}ski}, I., {Szewczyk}, O.,
  {Wyrzykowski}, {\L}., {Paczy{\'n}ski}, B., {Abe}, F., {Bond}, I.~A.,
  {Britton}, T.~R., {Gilmore}, A.~C., {Hearnshaw}, J.~B., {Itow}, Y., {Kamiya},
  K., {Kilmartin}, P.~M., {Korpela}, A.~V., {Masuda}, K., {Matsubara}, Y.,
  {Motomura}, M., {Muraki}, Y., {Nakamura}, S., {Okada}, C., {Ohnishi}, K.,
  {Rattenbury}, N.~J., {Sako}, T., {Sato}, S., {Sasaki}, M., {Sekiguchi}, T.,
  {Sullivan}, D.~J., {Tristram}, P.~J., {Yock}, P.~C.~M., {Yoshioka}, T., 2006.
  Discovery of a cool planet at 5.5 {E}arth masses through gravitational
  microlensing. Nature 439, 437--440.

\bibitem[{Belisheva et~al.(2005)Belisheva, Kuzhevskii, Vashenyuk, and
  Zhirov}]{Belisheva05}
Belisheva, N.~K., Kuzhevskii, B.~M., Vashenyuk, E.~V., Zhirov, V.~K., 2005.
  Correlation between the fusion dynamics of cells growing \textit{in vitro}
  and variations of neutron intensity near the {E}arth's surface. Doklady
  Biochemistry and Biophysics 402, 254--257.

\bibitem[{Buccino et~al.(2007)Buccino, Lemarchand, and Mauas}]{Buccino07}
Buccino, A.~P., Lemarchand, G.~A., Mauas, P. J.~D., 2007. {UV} habitable zones
  around {M} stars. Icarus 192, 582--587.

\bibitem[{Correia and Laskar(2003)}]{Correia03a}
Correia, A. C.~M., Laskar, J., 2003. Different tidal torques on a planet with a
  dense atmosphere and consequences to the spin dynamics. J. Geophys. Res.
  108~(E11), 9--1---9--10.

\bibitem[{Crutzen(1970)}]{Crutzen70}
Crutzen, P.~J., 1970. The influence of nitrogen oxides on the atmospheric ozone
  content. Q. J. R. Meteorol. Soc. 96, 320--325.

\bibitem[{Fichtner(2001)}]{Fichtner01}
Fichtner, H., 2001. Anomalous cosmic rays: {M}essengers from the outer
  heliosphere. Space Sci. Rev. 95, 639--754.

\bibitem[{Fridlund(2004)}]{Fridlund04}
Fridlund, C. V.~M., 2004. The {D}arwin mission. Adv. Space Res. 34, 613--617.

\bibitem[{Goldreich and Soter(1966)}]{Goldreich66}
Goldreich, P., Soter, S., 1966. Q in the solar system. Icarus 5, 375--389.

\bibitem[{Grenfell et~al.(2007{\natexlab{a}})Grenfell, Grie{\ss}meier, Patzer,
  Rauer, Segura, Stadelmann, Stracke, Titz, and von Paris}]{Grenfell07AB}
Grenfell, J.~L., Grie{\ss}meier, J.-M., Patzer, B., Rauer, H., Segura, A.,
  Stadelmann, A., Stracke, B., Titz, R., von Paris, P., 2007{\natexlab{a}}.
  Biomarker response to galactic cosmic ray-induced {NO}$_x$ and the methane
  greenhouse effect in the atmosphere of an {E}arth-like planet orbiting an {M}
  dwarf star. Astrobiology 7~(1), 208--221.

\bibitem[{Grenfell et~al.(2008)Grenfell, Grie{\ss}meier, Patzer, Rauer,
  Stracke, and von Paris}]{Grenfell08}
Grenfell, J.~L., Grie{\ss}meier, J.-M., Patzer, B., Rauer, H., Stracke, B., von
  Paris, P., 2008. Response of atmospheric biomarkers and related molecules to
  {NO$_x$} generated by stellar cosmic rays for planets in the habitable zone
  of active {M}-dwarf stars, {s}ubmitted.

\bibitem[{Grenfell et~al.(2007{\natexlab{b}})Grenfell, Stracke, von Paris,
  Patzer, Titz, Segura, and Rauer}]{Grenfell07PSS}
Grenfell, J.~L., Stracke, B., von Paris, P., Patzer, B., Titz, R., Segura, A.,
  Rauer, H., 2007{\natexlab{b}}. The response of atmospheric chemistry on
  earthlike planets around {F}, {G} and {K} stars to small variations in
  orbital distance. Planet. Space Sci. 55, 661--671.

\bibitem[{Grie{\ss}meier(2006)}]{GriessmeierPHD06}
Grie{\ss}meier, J.-M., 2006. Aspects of the magnetosphere-stellar wind
  interaction of close-in extrasolar planets. Ph.D. thesis, Technische
  Universit\"{a}t Braunschweig, {I}SBN 3-936586-49-7, {C}opernicus-GmbH
  Katlenburg-Lindau, {U}RL: http://www.digibib.tu-bs.de/?docid=00013336.

\bibitem[{Grie{\ss}meier et~al.(2005{\natexlab{a}})Grie{\ss}meier, Stadelmann,
  Lammer, Belisheva, and Motschmann}]{GriessmeierESLAB05}
Grie{\ss}meier, J.-M., Stadelmann, A., Lammer, H., Belisheva, N., Motschmann,
  U., 2005{\natexlab{a}}. The impact of galactic cosmic rays on extrasolar
  {E}arth-like planets in close-in habitable zones. In: Favata, F.,
  Sanz-Forcada, J., Gim\'{e}nez, A., Battrick, B. (Eds.), Proc. 39th ESLAB
  Symposium. pp. 305--309, {E}SA SP-588.

\bibitem[{Grie{\ss}meier et~al.(2005{\natexlab{b}})Grie{\ss}meier, Stadelmann,
  Motschmann, Belisheva, Lammer, and Biernat}]{Griessmeier05a}
Grie{\ss}meier, J.-M., Stadelmann, A., Motschmann, U., Belisheva, N.~K.,
  Lammer, H., Biernat, H.~K., 2005{\natexlab{b}}. Cosmic ray impact on
  extrasolar {E}arth-like planets in close-in habitable zones. Astrobiology
  5~(5), 587--603.

\bibitem[{Grie{\ss}meier et~al.(2004)Grie{\ss}meier, Stadelmann, Penz, Lammer,
  Selsis, Ribas, Guinan, Motschmann, Biernat, and Weiss}]{Griessmeier04}
Grie{\ss}meier, J.-M., Stadelmann, A., Penz, T., Lammer, H., Selsis, F., Ribas,
  I., Guinan, E.~F., Motschmann, U., Biernat, H.~K., Weiss, W.~W., 2004. The
  effect of tidal locking on the magnetospheric and atmospheric evolution of
  ``{H}ot {J}upiters''. Astron. Astrophys. 425, 753--762.

\bibitem[{Grie{\ss}meier et~al.(2007)Grie{\ss}meier, Zarka, and
  Spreeuw}]{Griessmeier07AA}
Grie{\ss}meier, J.-M., Zarka, P., Spreeuw, H., 2007. Predicting low-frequency
  radio fluxes of known extrasolar planets. Astron. Astrophys. 475, 359--368.

\bibitem[{Haagen-Smit(1952)}]{HaagenSmit52}
Haagen-Smit, A., 1952. Chemistry and physiology of {L}os {A}ngeles smog.
  Indust. Eng. Chem. 44, 1342--1346.

\bibitem[{Heath et~al.(1999)Heath, Doyle, Joshi, and Haberle}]{Heath99}
Heath, M.~J., Doyle, L.~R., Joshi, M.~M., Haberle, R.~M., 1999. Habitability of
  planets around red dwarf stars. Origins Life Evol. Biosphere 29, 405--424.

\bibitem[{Heber et~al.(2006)Heber, Fichtner, and Scherer}]{Heber06}
Heber, B., Fichtner, H., Scherer, K., 2006. Solar and heliospheric modulation
  of galactic cosmic rays. Space Sci. Rev. 125, 81--93.

\bibitem[{Holzwarth and Jardine(2007)}]{Holzwarth07}
Holzwarth, V., Jardine, M., 2007. Theoretical mass loss rates of cool
  main-sequence stars. Astron. Astrophys. 463, 11--21.

\bibitem[{Hubbard(1984)}]{Hubbard84}
Hubbard, W.~B., 1984. Planetary interiors. Van Nostrand Reinhold Co., New York.

\bibitem[{Joshi(2003)}]{Joshi03}
Joshi, M., 2003. Climate model studies of synchronously rotating planets.
  Astrobiology 3~(2), 415--427.

\bibitem[{Joshi et~al.(1997)Joshi, Haberle, and Reynolds}]{Joshi97}
Joshi, M.~M., Haberle, R.~M., Reynolds, R.~T., 1997. Simulations of the
  atmospheres of synchronously rotating terrestrial planets orbiting {M}
  dwarfs: Conditions for atmospheric collapse and the implications for
  habitability. Icarus 129, 450--465.

\bibitem[{Kallenrode(2000)}]{Kallenrode00}
Kallenrode, M.-B., 2000. Galactic cosmic rays. In: Scherer, K., Fichtner, H.,
  Marsch, E. (Eds.), The Outer Heliosphere: Beyond the Planets. Copernicus
  Ges., Katlenburg-Lindau, Ch.~7, pp. 165--190.

\bibitem[{Kasting et~al.(1993)Kasting, Whitmire, and Reynolds}]{Kasting93}
Kasting, J.~F., Whitmire, D.~P., Reynolds, R.~T., 1993. Habitable zones around
  main sequence stars. Icarus 101, 108--128.

\bibitem[{Khodachenko et~al.(2007)Khodachenko, Ribas, Lammer, Grie{\ss}meier,
  Leitner, Selsis, Eiroa, Hanslmeier, Biernat, Farrugia, and
  Rucker}]{Khodachenko07AB}
Khodachenko, M.~L., Ribas, I., Lammer, H., Grie{\ss}meier, J.-M., Leitner, M.,
  Selsis, F., Eiroa, C., Hanslmeier, A., Biernat, H.~K., Farrugia, C.~J.,
  Rucker, H.~O., 2007. Coronal {M}ass {E}jection ({CME}) activity of low mass
  {M} stars as an important factor for the habitability of terrestrial
  exoplanets. {I}. {CME} impact on expected magnetospheres of {E}arth-like
  exoplanets in close-in habitable zones. Astrobiology 7~(1), 167--184.

\bibitem[{Kuchner(2003)}]{Kuchner03}
Kuchner, M.~J., 2003. Volatile-rich {E}arth-mass planets in the habitable zone.
  Astrophys. J. 596, L105--L108.

\bibitem[{Lammer et~al.(2008)Lammer, Khodachenko, Lichtenegger, and
  Kulikov}]{Lammer08}
Lammer, H., Khodachenko, M.~L., Lichtenegger, H. I.~M., Kulikov, Y.~N., 2008.
  Impact of stellar activity on the evolution of planetary atmospheres and
  habitability. In: Dvorak, R. (Ed.), Extrasolar Planets: Formation, Detection
  and Dynamics. WILEY-VCH Verlag.

\bibitem[{Lammer et~al.(2007)Lammer, Lichtenegger, Kulikov, Grie{\ss}meier,
  Terada, Erkaev, Biernat, Khodachenko, Ribas, Penz, and Selsis}]{Lammer07AB}
Lammer, H., Lichtenegger, H. I.~M., Kulikov, Y.~N., Grie{\ss}meier, J.-M.,
  Terada, N., Erkaev, N.~V., Biernat, H.~K., Khodachenko, M.~L., Ribas, I.,
  Penz, T., Selsis, F., 2007. Coronal {M}ass {E}jection ({CME}) activity of low
  mass {M} stars as an important factor for the habitability of terrestrial
  exoplanets. {II}. {CME} induced ion pick up of {E}arth-like exoplanets in
  close-in habitable zones. Astrobiology 7~(1), 185--207.

\bibitem[{L\'{e}ger et~al.(2004)L\'{e}ger, Selsis, Sotin, Guillot, Despois,
  Mawet, Ollivier, Lab\`{e}que, Brachet, Chazelas, and Lammer}]{Leger04}
L\'{e}ger, A., Selsis, F., Sotin, C., Guillot, T., Despois, D., Mawet, D.,
  Ollivier, M., Lab\`{e}que, F.~A., Brachet, C.~V., Chazelas, B., Lammer, H.,
  2004. A new family of planets? ``{O}cean-{P}lanets''. Icarus 169, 499--504.

\bibitem[{Lestrade et~al.(2006)Lestrade, Wyatt, Bertoldi, Dent, and
  Menten}]{Lestrade06}
Lestrade, J.-F., Wyatt, M.~C., Bertoldi, F., Dent, W. R.~F., Menten, K.~M.,
  2006. Search for cold debris disks around {M}-dwarfs. Astron. Astrophys. 460,
  733--741.

\bibitem[{Lovelock(1965)}]{Lovelock65}
Lovelock, J.~E., 1965. A physical basis for life detection experiments. Nature
  207, 568--670.

\bibitem[{{Lovis} et~al.(2006){Lovis}, {Mayor}, {Pepe}, {Alibert}, {Benz},
  {Bouchy}, {Correia}, {Laskar}, {Mordasini}, {Queloz}, {Santos}, {Udry},
  {Bertaux}, and {Sivan}}]{Lovis06}
{Lovis}, C., {Mayor}, M., {Pepe}, F., {Alibert}, Y., {Benz}, W., {Bouchy}, F.,
  {Correia}, A.~C.~M., {Laskar}, J., {Mordasini}, C., {Queloz}, D., {Santos},
  N.~C., {Udry}, S., {Bertaux}, J.-L., {Sivan}, J.-P., 2006. An extrasolar
  planetary system with three {N}eptune-mass planets. Nature 441, 305--309.

\bibitem[{MacDonald(1964)}]{MacDonald64}
MacDonald, G. J.~F., 1964. Tidal friction. Rev. Geophys. 2~(3), 467--541.

\bibitem[{Mann et~al.(1999)Mann, Jansen, MacDowall, Kaiser, and Stone}]{Mann99}
Mann, G., Jansen, F., MacDowall, R.~J., Kaiser, M.~L., Stone, R.~G., 1999. A
  heliospheric model and type {III} radio bursts. Astron. Astrophys. 348,
  614--620.

\bibitem[{McDonald et~al.(1992)McDonald, Moraal, Reinecke, Lal, and
  McGuire}]{McDonald92}
McDonald, F.~B., Moraal, H., Reinecke, J. P.~L., Lal, N., McGuire, R.~E., 1992.
  The cosmic radiation in the heliosphere at successive solar minima. J.
  Geophys. Res. 97~(A2), 1557--1570.

\bibitem[{Murray and Dermott(1999)}]{Murray99}
Murray, C.~D., Dermott, S.~F., 1999. Solar System Dynamics. Cambridge
  University Press, Cambridge.

\bibitem[{Parker(1958)}]{Parker58}
Parker, E.~N., 1958. Dynamics of the interplanetary gas and magnetic fields.
  Astrophys. J. 128, 664--676.

\bibitem[{Peale(1999)}]{Peale99}
Peale, S.~J., 1999. Origin and evolution of the natural satellites. Ann. Rev.
  Astron. Astrophys. 37, 533--602.

\bibitem[{Pr\"{o}lss(2004)}]{Proelss04}
Pr\"{o}lss, G.~W., 2004. Physics of the Earth's Space Environment.
  Springer-Verlag, Berlin.

\bibitem[{Reeves et~al.(1992)Reeves, Cayton, Gary, and Belian}]{Reeves92}
Reeves, G.~D., Cayton, T.~E., Gary, S.~P., Belian, R.~D., 1992. The great solar
  energetic particle events of 1989 observed from geosynchronous orbit. J.
  Geophys. Res. 97~(A5), 6219--6226.

\bibitem[{Ribas et~al.(2008)Ribas, Font-Ribera, and Beaulieu}]{Ribas08}
Ribas, I., Font-Ribera, A., Beaulieu, J.-P., 2008. A $\sim5 {M}_\oplus$
  super-{E}arth orbiting {GJ} 436? {T}he power of near-grazing transits.
  Astrophys. J. 677, L59--L62.

\bibitem[{Rivera et~al.(2005)Rivera, Lissauer, Butler, Marcy, Vogt, Fischer,
  Brown, Laughlin, and Henry}]{Rivera05}
Rivera, E.~J., Lissauer, J.~J., Butler, R.~P., Marcy, G.~W., Vogt, S.~S.,
  Fischer, D.~A., Brown, T.~M., Laughlin, G., Henry, G.~W., 2005. A $\sim 7.5$
  ${M}_\oplus$ planet orbiting the nearby star, {GJ} 876. Astrophys. J. 634,
  625--640.

\bibitem[{Sagan et~al.(1993)Sagan, Thomson, Carlson, Gurnett, and
  Hord}]{Sagan93}
Sagan, C., Thomson, W.~R., Carlson, R., Gurnett, D., Hord, C., 1993. A search
  for life on earth from the Galileo spacecraft. Nature 365, 715--721.

\bibitem[{Scalo et~al.(2007)Scalo, Kaltenegger, Segura, Fridlund, Ribas,
  Kulikov, Grenfell, Rauer, Odert, Leitzinger, Selsis, Khodachenko, Eiroa,
  Kasting, and Lammer}]{Scalo07}
Scalo, J., Kaltenegger, L., Segura, A., Fridlund, M., Ribas, I., Kulikov,
  Y.~N., Grenfell, J.~L., Rauer, H., Odert, P., Leitzinger, M., Selsis, F.,
  Khodachenko, M.~L., Eiroa, C., Kasting, J., Lammer, H., 2007. M stars as
  targets for terrestrial exoplanet searches and biosignature detection.
  Astrobiology 7~(1), 85--166.

\bibitem[{Scherer et~al.(2008)Scherer, Fichtner, Heber, Ferreia, and
  Potgieter}]{Scherer08}
Scherer, K., Fichtner, H., Heber, B., Ferreia, S. E.~S., Potgieter, M.~S.,
  2008. Cosmic ray flux at the {E}arth in a variable heliosphere. Adv. Space
  Res. 41, 1171--1176.

\bibitem[{Schindler and Kasting(2000)}]{Schindler00}
Schindler, T.~L., Kasting, J.~F., 2000. Synthetic spectra of simulated
  terrestrial atmospheres containing possible biomarker gases. Icarus 145,
  262--271.

\bibitem[{Schneider et~al.(2006)Schneider, Riaud, Tinetti, Schmid, Stam, Udry,
  Baudoz, Boccaletti, Grasset, Mawet, Surdej, and {The See-Coast
  Team}}]{Schneider06}
Schneider, J., Riaud, P., Tinetti, G., Schmid, H.~M., Stam, D., Udry, S.,
  Baudoz, P., Boccaletti, A., Grasset, O., Mawet, D., Surdej, J., {The
  See-Coast Team}, 2006. {SEE-COAST}: The super-{E}arth explorer. In: Barret,
  D., Casoli, F., Contini, T., Lagache, G., Levacelier, A., Pagani, L. (Eds.),
  SF2A-2006: Semaine de l'Astrophysique Francaise. pp. 429--432.

\bibitem[{Selsis(2004)}]{Selsis04}
Selsis, F., 2004. The atmosphere of terrestrial exoplanets: Detection and
  characterization. In: Beaulieu, J.-P., {Lecavelier des Etangs}, A., Terquem,
  C. (Eds.), ASP Conf. Ser. 321: Extrasolar Planets: Today and Tomorrow. pp.
  170--181.

\bibitem[{Selsis et~al.(2007)Selsis, Chazelas, Bord\'{e}, Ollivier, Brachet,
  Decaudin, Bouchy, Ehrenreich, Grie{\ss}meier, Lammer, Sotin, Grasset, Moutou,
  Barge, Deleuil, Mawet, Despois, Kasting, and L\'{e}ger}]{Selsis07}
Selsis, F., Chazelas, B., Bord\'{e}, P., Ollivier, M., Brachet, F., Decaudin,
  M., Bouchy, F., Ehrenreich, D., Grie{\ss}meier, J.-M., Lammer, H., Sotin, C.,
  Grasset, O., Moutou, C., Barge, P., Deleuil, M., Mawet, D., Despois, D.,
  Kasting, J.~F., L\'{e}ger, A., 2007. Could we identify hot ocean-planets with
  {CoRoT}, {K}epler and {D}oppler velocimetry? Icarus 191, 453--468.

\bibitem[{Selsis et~al.(2002)Selsis, Despois, and Parisot}]{Selsis02}
Selsis, F., Despois, D., Parisot, J.-P., 2002. Signature of life on exoplanets:
  Can {D}arwin produce false positive detections? Astron. Astrophys. 388,
  985--1003.

\bibitem[{Seo et~al.(1994)Seo, McDonald, Lal, and Webber}]{Seo94}
Seo, E.~S., McDonald, F.~B., Lal, N., Webber, W.~R., 1994. Study of cosmic-ray
  {H} and {He} isotopes at 23 {AU}. Astrophys. J. 432, 656--664.

\bibitem[{Shea and Smart(2000)}]{Shea00}
Shea, M.~A., Smart, D.~F., 2000. Cosmic ray implications for human health.
  Space Sci. Rev. 93, 187--205.

\bibitem[{Smart et~al.(2000)Smart, Shea, and Fl\"{u}ckiger}]{Smart00}
Smart, D.~F., Shea, M.~A., Fl\"{u}ckiger, E.~O., 2000. Magnetospheric models
  and trajectory computations. Space Sci. Rev. 93, 305--333.

\bibitem[{Smith et~al.(2004)Smith, Scalo, and Wheeler}]{Smith04}
Smith, D.~S., Scalo, J., Wheeler, J.~C., 2004. Transport of ionising radiation
  in terrestrial-like exoplanet atmospheres. Icarus 171, 229--253.

\bibitem[{Stadelmann(2005)}]{Stadelmann04}
Stadelmann, A., 2005. Globale {E}ffekte einer {E}rdmagnetfeldumkehrung:
  {M}agnetosph\"{a}renstruktur und kosmische {T}eilchen. Ph.D. thesis,
  Technische Universit\"{a}t Braunschweig, {I}SBN 3-936586-42-{X},
  {C}opernicus-GmbH Katlenburg-Lindau, {U}RL:
  http://www.digibib.tu-bs.de/?docid=00000002.

\bibitem[{{Tarter} et~al.(2007){Tarter}, {Backus}, {Mancinelli}, {Aurnou},
  {Backman}, {Basri}, {Boss}, {Clarke}, {Deming}, {Doyle}, {Feigelson},
  {Freund}, {Grinspoon}, {Haberle}, {Hauck}, {Heath}, {Henry}, {Hollingsworth},
  {Joshi}, {Kilston}, {Liu}, {Meikle}, {Reid}, {Rothschild}, {Scalo}, {Segura},
  {Tang}, {Tiedje}, {Turnbull}, {Walkowicz}, {Weber}, and {Young}}]{Tarter07}
{Tarter}, J.~C., {Backus}, P.~R., {Mancinelli}, R.~L., {Aurnou}, J.~M.,
  {Backman}, D.~E., {Basri}, G.~S., {Boss}, A.~P., {Clarke}, A., {Deming}, D.,
  {Doyle}, L.~R., {Feigelson}, E.~D., {Freund}, F., {Grinspoon}, D.~H.,
  {Haberle}, R.~M., {Hauck}, II, S.~A., {Heath}, M.~J., {Henry}, T.~J.,
  {Hollingsworth}, J.~L., {Joshi}, M.~M., {Kilston}, S., {Liu}, M.~C.,
  {Meikle}, E., {Reid}, I.~N., {Rothschild}, L.~J., {Scalo}, J., {Segura}, A.,
  {Tang}, C.~M., {Tiedje}, J.~M., {Turnbull}, M.~C., {Walkowicz}, L.~M.,
  {Weber}, A.~L., {Young}, R.~E., 2007. A reappraisal of the habitability of
  planets around {M} dwarf stars. Astrobiology 7~(1), 30--65.

\bibitem[{Udry et~al.(2007)Udry, Bonfils, Delfosse, Forveille, Mayor, Perrier,
  Bouchy, Lovis, Pepe, Queloz, and Bertaux}]{Udry07}
Udry, S., Bonfils, X., Delfosse, X., Forveille, T., Mayor, M., Perrier, C.,
  Bouchy, F., Lovis, C., Pepe, F., Queloz, D., Bertaux, J.-L., 2007. The
  {HARPS} search for southern extra-solar planets. {XI}. {S}uper-{E}arths (5
  and 8 ${M}_\oplus$) in a 3-planet system. Astron. Astrophys. 469, L43--L47.

\bibitem[{Valencia et~al.(2006)Valencia, O'Connell, and Sasselov}]{Valencia06}
Valencia, D., O'Connell, R.~J., Sasselov, D., 2006. Internal structure of
  massive terrestrial planets. Icarus 181, 545--554.

\bibitem[{Valencia et~al.(2007)Valencia, Sasselov, and O'Connell}]{Valencia07b}
Valencia, D., Sasselov, D.~D., O'Connell, R.~J., 2007. Detailed models of
  super-{E}arths: How well can we infer bulk properties? Astrophys. J. 665,
  1413--1420.

\bibitem[{Vogt et~al.(2007)Vogt, Zieger, Glassmeier, Stadelmann, Kallenrode,
Sinnhuber and Winkler}]{Vogt07}
Vogt, J., Zieger, B., Glassmeier, K.-H., Stadelmann, A., 
   Kallenrode, M.-B., Sinnhuber, M., Winkler, H., 2007. 
   Energetic particles in the paleomagnetosphere: {R}educed dipole
   configurations and quadrupolar contributions. J. Geophys. Res. 112 (A11).
   A06216.

\bibitem[{Voigt(1981)}]{Voigt81}
Voigt, G.-H., 1981. A mathematical magnetospheric field model with independent
  physical parameters. Planet. Space Sci. 29, 1--20.

\bibitem[{Voigt(1995)}]{Voigt95}
Voigt, G.-H., 1995. Magnetospheric configuration. In: Volland, H. (Ed.),
  Handbook of atmospheric electrodynamics. Vol.~II. CRC Press, Ch.~11, pp.
  333--388.

\bibitem[{Wood et~al.(2002)Wood, M\"{u}ller, Zank, and Linsky}]{Wood02}
Wood, B.~E., M\"{u}ller, H.-R., Zank, G.~P., Linsky, J.~L., 2002. Measured
  mass-loss rates of solar-like stars as a function of age and activity.
  Astrophys. J. 574, 412--425.

\bibitem[{Wood et~al.(2005)Wood, M\"{u}ller, Zank, Linsky, and
  Redfield}]{Wood05}
Wood, B.~E., M\"{u}ller, H.-R., Zank, G.~P., Linsky, J.~L., Redfield, S., 2005.
  New mass-loss measurements from astrospheric {L}y$\alpha$ absorption.
  Astrophys. J. 628, L143--L146.

\end{thebibliography}
\end{document}